\documentclass[a4paper]{article}

\usepackage{INTERSPEECH2016}

\usepackage{graphicx}
\usepackage{amssymb,amsmath,bm}
\usepackage{textcomp}

\ninept
\title{Statistical Parametric Speech Synthesis Using Bottleneck Representation From Sequence Auto-encoder}

\makeatletter
\def\name#1{\gdef\@name{#1\\}}
\makeatother \name{{\em Sivanand Achanta, KNRK Raju Alluri and Suryakanth V Gangashetty}}

\address{Speech and Vision Laboratory, IIIT Hyderabad, INDIA \\
  {\small \tt \{sivanand.a,raju.alluri\}@research.iiit.ac.in, svg@iiit.ac.in}
}

\begin{document}

  \maketitle
  \begin{abstract}
    In this paper, we describe a statistical parametric speech synthesis approach with unit-level acoustic representation. In conventional deep neural network based speech synthesis, the input text features are repeated for the entire duration of phoneme for mapping text and speech parameters. This mapping is learnt at the frame-level which is the de-facto acoustic representation. However much of this computational requirement can be drastically reduced if every unit can be represented with a fixed-dimensional representation. Using recurrent neural network based auto-encoder, we show that it is indeed possible to map units of varying duration to a single vector. We then use this acoustic representation at unit-level to synthesize speech using deep neural network based statistical parametric speech synthesis technique. Results show that the proposed approach is able to synthesize at the same quality as the conventional frame based approach at a highly reduced computational cost.    
  \end{abstract}
  \noindent{\bf Index Terms}: speech synthesis, encoder-decoder, recurrent neural network

  \section{Introduction}

    There are three fundamental problems in statistical parametric speech synthesis (SPSS) that have been outlined \cite{sps}, one of them is acoustic modeling. Traditionally hidden Markov models (HMM) were used for SPSS \cite{hmmspss}, however there is now growing interest in the community to use deep learning techniques for acoustic modeling in SPSS. It has been demonstrated in \cite{dnnsps} \cite{dnnsps2} that deep neural networks (DNN) can perform better than traditional HMMs for SPSS. Recently, DNNs have been replaced by recurrent neural networks (RNNs) \cite{ulstmspss} \cite{blstmspss} \cite{sivarnn} as they better capture the temporal dependencies.
    
    However both these models use traditional frame-level representation of speech signals for mapping text features to speech parameters. One draw back of this is that the text features have to be repeated and made to evolve at the same frame-rate as speech parameters which increases the computation. This can be reduced drastically if the speech parameters of a unit (for the experiments in this paper unit is chosen as phoneme and hence terms unit and phoneme are interchangeably used) are given a fixed dimensional representation and then mapping takes place at phoneme level. This manner of statistical mapping of features at unit-level has remained unexplored so far within the SPSS community. Since the duration of different units vary, the fundamental problem is to find a mechanism for representing these units with a single or fixed number of multiple vectors.  
    
    Thanks to the recent sequence-to-sequence learning techniques introduced in \cite{graves2012sequence} \cite{sutseq} which allow us to map sequences of varying lengths. Typically this is achieved via neural network encoder-decoder architectures \cite{cholearning}. Broadly, the encoder-decoder models can be grouped into two types for mapping one sequence to another sequence of different lengths, one is to map the input sequence to a fixed dimensional vector and then to map it to the target sequence. We refer to this architecture as encoder-decoder with fixed vector context representation (ED-FVC) throughout this paper \cite{sutseq} \cite{cholearning}. The other is to soft-search the input features that align with the output using an alignment model \cite{dbnmt}. 
    
    The first contribution of this paper is that, we explore RNN encoder-decoder for auto-encoding the acoustic features of units to a single vector referred to as recurrent bottleneck feature (RBN). Auto-encoders using RNN encoder-decoder architecture have earlier been used for semi-supervised sequence learning and as pre-training in \cite{dai2015semi} \cite{danmlpb}. However, the RBN feature, which is exploited in this work, has never been directly used in the prior works. This method detailed in section \ref{sec:ED-FVC} can be viewed as RNN based analysis-by-synthesis (AbS) of speech using the explicit unit-boundary information. 
    
    Once we train the auto-encoder, these RBN features can be used for mapping from text-features. In our experiments DNN is used for the purpose of mapping text to unit-level acoustic features, although, in principle any other regression technique like classification and regression trees \cite{breiman1984classification} or random forests \cite{breiman2001random} \cite{clustergen} \cite{rfspss} can be used. In addition, these RBN features can be used in unit-selection based approaches for computing the concatenation cost and/or for the final synthesis.
   
   The second contribution of this paper is that, simple recurrent neural (SRN) units are used as basic RNN units as opposed to the typically used long-short term memory units (LSTM) \cite{lstm} which have complicated gating mechanisms. In the RNN encoder-decoder network, using SRNs as units we show that it is indeed possible to auto-encode units. Although SRNs have been proposed as an alternative to LSTMs, they have not been used as part of sequence-to-sequence neural networks so far. 
   
The paper is organized as follows: In section \ref{sec:ED-FVC} we describe the architecture of the RNN encoder-decoder model which enables us to auto-encode units, section \ref{sec:exp} describes AbS and SPSS experiments using RBN features and its results. The conclusions and scope for future work are finally presented in section \ref{sec:con}.

\begin{figure*}[htb]
\begin{minipage}[b]{1.0\linewidth}
\centering
\vspace{-1cm}
\includegraphics[scale=0.4]{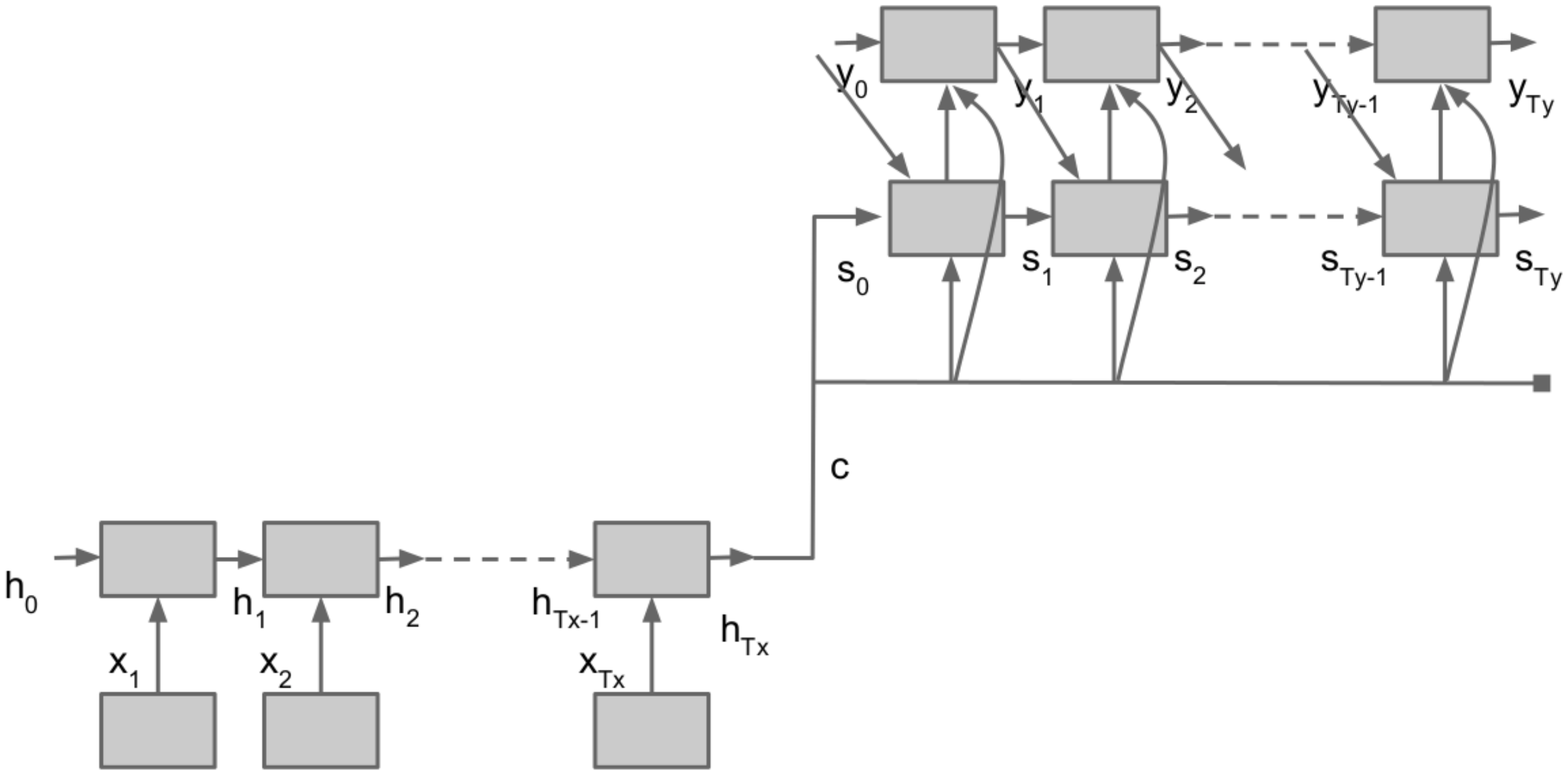}
\vspace{-2cm}
\caption{Encoder-Decoder architecture with fixed vector representation for context vector (ED-FVC), x$_t$, h$_t$, c, s$_t$, \text{ and }y$_t$ represent input,encoder hidden state, context vector, decoder hidden state, and output at time $t$.}
\label{ED-FVC} 
\end{minipage}
\end{figure*}

  
\section{RNN Encoder-Decoder Model}
\label{sec:ED-FVC}
In this section we describe the architecture of encoder-decoder network with fixed vector context.

\subsection{ED-FVC}
The block diagram of ED-FVC is presented in Fig. \ref{ED-FVC}. ED-FVC architecture consists of a uni-directional RNN as encoder and decoder. The last hidden state of encoder is taken as initial hidden state of the decoder and also as the context vector. It is this last hidden state of the encoder which we refer to as RBN. 

\subsubsection {SRN}
In this section, we briefly review SRNs which will be used in the ED-FVC network. Typically the encoder-decoder architectures use either LSTMs \cite{sutseq} or gated-recurrent units (GRU) \cite{cho2014properties} \cite{dbnmt} \cite{cholearning} as RNN units, to avoid the vanishing gradients problem \cite{bengio1994learning}. However, in this paper we use simple RNNs. There are two reasons why SRNs can be preferred over LSTMs/GRUs: (1) The number of parameters in LSTM/GRU are far more for a given hidden state size compared to simple RNN and (2) It is difficult to know which of those gates are actually contributing to the performance. Recently, there have been some works which have tried to address the latter drawback \cite{lstmsso} \cite{eernn}, and specifically in the context of SPSS in \cite{wu2016investigating}.

SRNs are RNNs without any advanced gating mechanisms. It has been shown in previous studies that Elman RNNs \cite{elman} work well if the initialization scheme used is more robust and gradient clipping is employed to avoid gradients overflow \cite{advrnn} \cite{le2015simple} \cite{nprnn}. In \cite{le2015simple}, authors proposed to solve the vanishing gradient problem using diagonal initialization which was inspired by orthogonal initialization suggested in \cite{saxeexact} and sparse initialization. More recently \cite{nprnn} reports SRNs being able to memorize long-term dependencies by initializing recurrent weight matrix with a special structure and using rectified linear units \cite{nairrelu}. However, these have not been explored in the context of sequence to sequence learning. In this work we propose to use IRNN (SRNs with recurrent weight matrix initialized to identity) in the encoder-decoder models. Also we note that authors in \cite{le2015simple} suggest ReLU units with the identity initialization. However, we in our experiments found that gradients explode quickly when using ReLU units, instead we use \textit{tanh} nonlinearity for our experiments.

\subsubsection{RNN Encoder}
The encoder in our case is a uni-directional recurrent neural network. The encoder reads the entire input sequence and a representation is stored in the final state vector. 
\begin{equation}
\begin{split}
  h_{ft} = ~ & f(W_{fi}x_t + W_fh_{t-1})
  \label{fp}
  \end{split}
\end{equation}
$W_{fi}, W_f$ are the input and recurrent weight matrices corresponding to recurrent layer.
%

\subsubsection{Fixed-dimensional context vector}
Context vector is formed by taking the final state vector of encoder. This context vector is used as an initial state vector $s_0$ in the decoder similar to \cite{sutseq}
\begin{equation}
\begin{split}
  c = ~ & [h_{fT_x}]'\\
  s_0 = ~ & c
\end{split}
\end{equation}

\subsubsection{RNN Decoder}
The decoder is also a uni-directional RNN.
\begin{equation}
\begin{split}
  s_t = ~ & f(W_iy_{t-1} + Ws_{t-1} + W_cc)\\
  y_t = ~ & g(U_is_t + Uy_{t-1} + U_cc)
  \label{RNN_Dec}
  \end{split}
\end{equation}
$W_i,W$ and $W_c$ denote the weight connections from past output, past hidden state and context layers. It has to be noted that the past output $y_{t-1}$, during training, can be either from the ground truth or from the prediction of the network itself. Similarly, $U_i,U$ and $U_c$ denote the weight connections from current hidden state, past output and context layers. $g$ in our case is a linear layer. 
 
\subsubsection{Back-propagation of error signal}
The following equations describe the calculation of error signal for updating the weights of encoder-decoder structure. The computation of error signal for decoder is given as:

\begin{equation}
\begin{split}
  \delta_{y_t} = ~ & D_{g_t}(W_i\delta_{s_{t+1}} + U\delta_{y_{t+1}} + i_{y_t})\\
  i_{s_t} = ~ & U_i\delta_{y_t}\\
  \delta_{s_t} = ~ & D_{f_t}(W\delta_{s_{t+1}} + i_{s_t})\\  
  \label{bp_dec}
  \end{split}
\end{equation}

where $i_{y_t}$ is the injected error signal which for our choice of $g$ and squared error loss function is $y_t - d_t$, $d_t$ is the target output. $D_{g_t},D_{f_t}$ are the derivatives of the functions at output and hidden layers respectively.

The error signal for encoder are computed
\begin{equation}
\begin{split}
  \delta_{h_{ft}} = ~ & D_{h_{ft}}(W_f\delta_{h_{f(t+1)}} + i_{h_{ft}})\\
  \label{bp_enc}
  \end{split}
\end{equation}

The injected error signal for the encoder is non-zero only at the final state $T_x$ at all other times it is zero and $D_{h_{ft}}$ is the derivative of the function at hidden layer. This error signal is then back-propagated recursively from $T_x$ all the way down to initial time instant.

In the next section we describe experiments using the ED-FVC architecture as an auto-encoder. Since the input and output to the ED-FVC model are same we refer to this model as recurrent auto-encoder (RAE) from here on.

\begin{figure}[htb]
\begin{minipage}[b]{1\linewidth}
\centering
\includegraphics[scale=0.5]{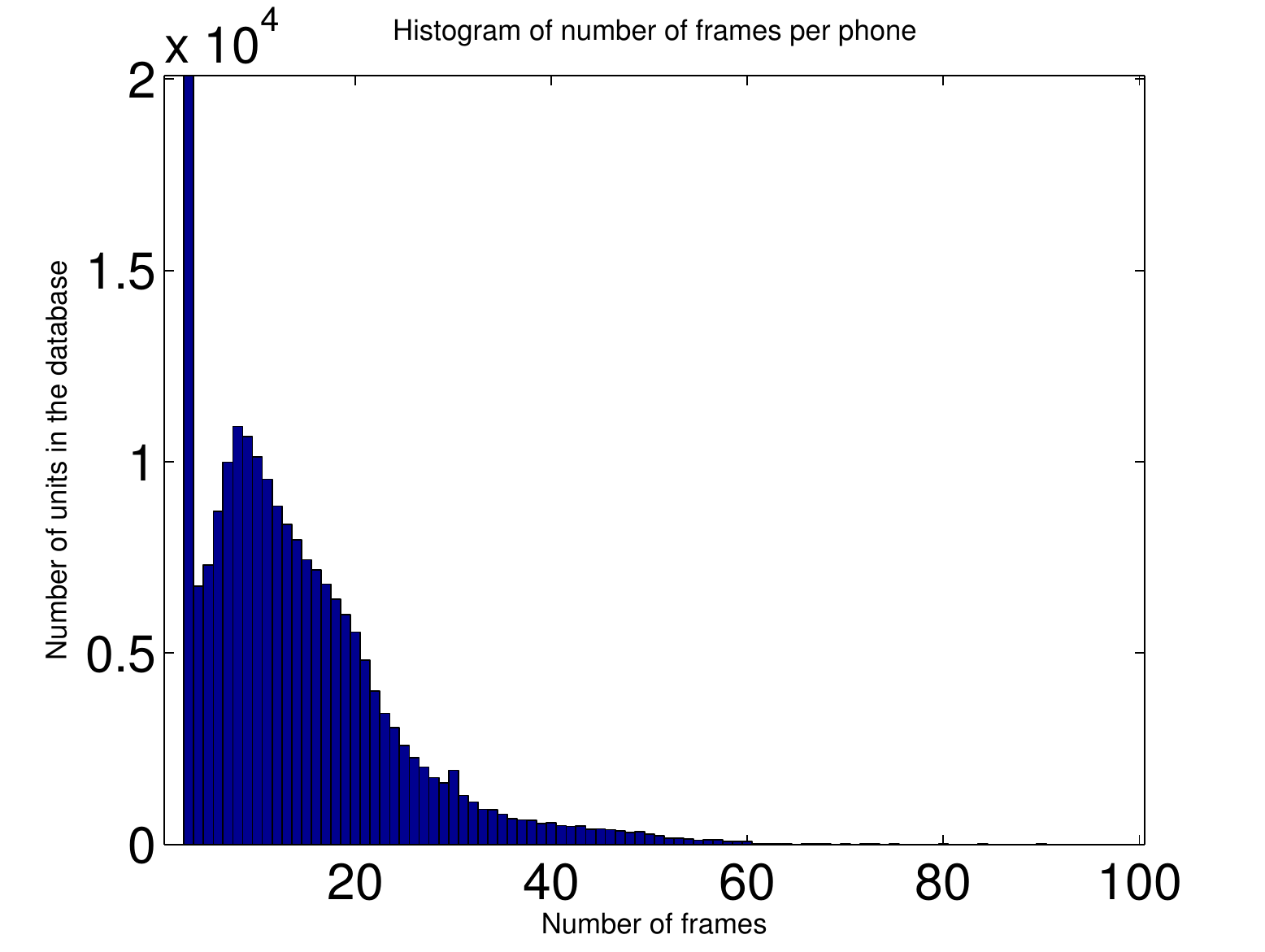}
\caption{Histogram of number of frames per phone}
\label{fig:histnofp} 
\end{minipage}
\end{figure}

\begin{figure}[htb]
\begin{minipage}[b]{1.0\linewidth}
\centering
\vspace{-1cm}
\includegraphics[scale=0.4]{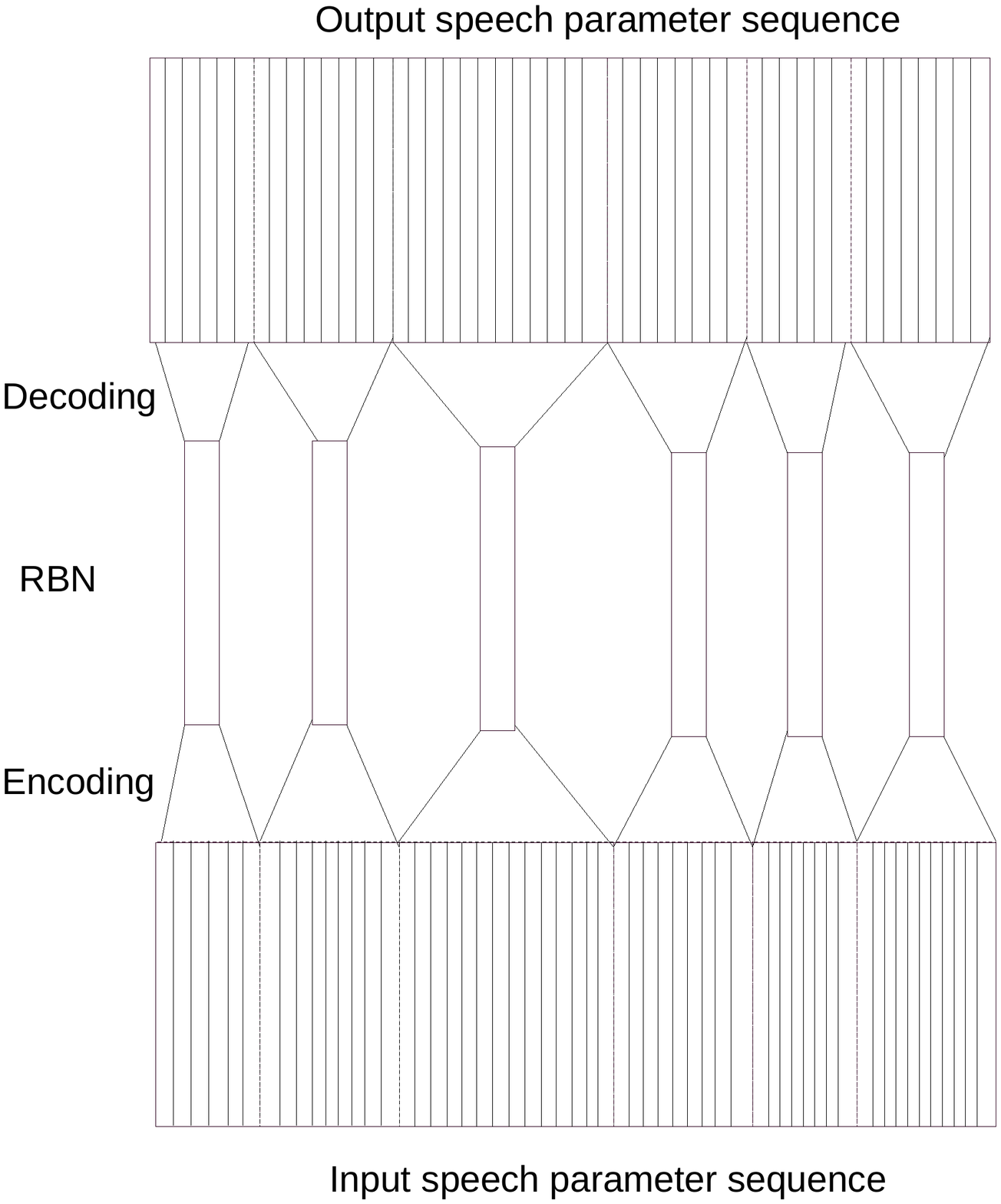}
\vspace{-2cm}
\caption{Schematic of proposed AbS using RBN feature}
\label{fig:RBN} 
\end{minipage}
\end{figure}

  \section{Experiments and Results}
  \label{sec:exp}
  We use a subset of Blizzard challenge 2015 database for our experiments. Blizzard challenge 2015 database contains about 4 hours of speech data in each of three Indian languages (Hindi, Tamil and Telugu), and about 2 hours of speech data in each of other three Indian languages (Marathi, Bengali and Malayalam), all recorded by native professional speakers in high quality studio environments. We used Telugu language dataset for our experiments. The speech recordings released were sampled at 16KHz. Phone-level alignments were performed using the EHMM tool \cite{prahallad2010automatic}.
  
  50 dimensional Mel-general cepstral (MGC) features and 26 dimensional band-aperiodicities (BAP) were extracted with a frame-shift of 5 ms for all the speech utterances along with their deltas and double-deltas. This feature extraction is followed from HTS-STRAIGHT demo available online. In all our experiments, natural $f_0$, BAP (when not predicted from network) and duration were used during synthesis. There are in total around 200000 units in the dataset which was divided into training/validation/test sets as 188000, 5660 and 5660 units respectively. The duration of units typically ranged from 4 to 30 frames as can be seen from Fig. \ref{fig:histnofp}. 
    
    \begin{figure*}[htb]
\begin{minipage}[b]{1.0\linewidth}
\centering
\includegraphics[scale=0.5]{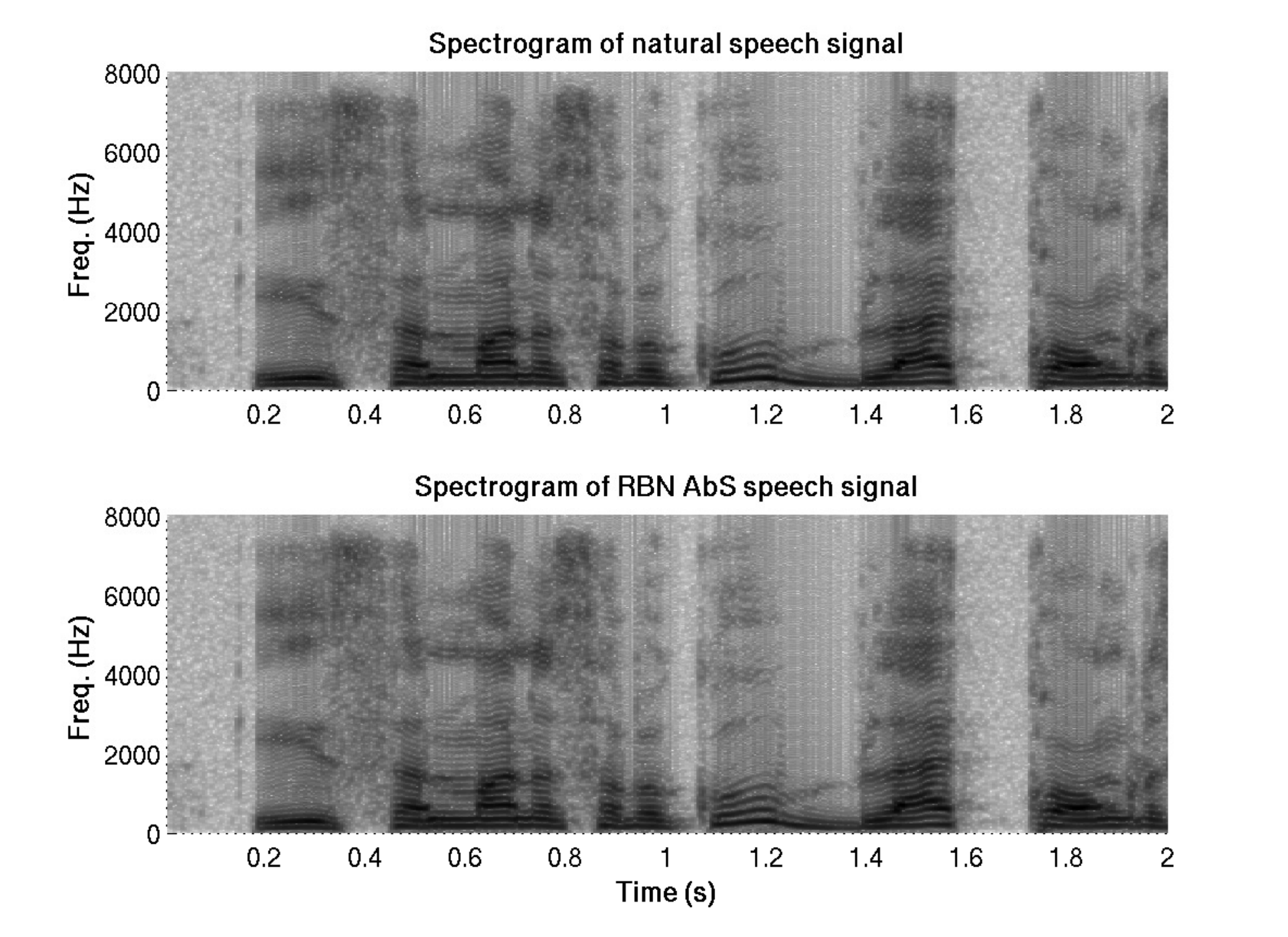}
\caption{Spectrograms of natural and RBN AbS speech signals}
\label{fig:RBNAbS_sgram} 
\end{minipage}
\end{figure*}

  \subsection{RAE based AbS experiments}
   
  Based on the labels, parametric sequences of each phone are prepared for training the RAE. The architecture of RAE used was enc = \{xL 500N\} , dec = \{500N xL\}, where ``L" represents linear units and ``N" represents $tanh$ units. 
  
  The training process is schematically represented in Fig. \ref{fig:RBN}, each unit along with its boundaries demarcated by dashed-lines are presented frame-wise (frames indicated with solid lines) to the RAE during training. The mean squared error over the entire training set is minimized. Once the training is complete, RBN features are obtained by presenting the units to the encoder and extracting the last hidden state. These RBN features along with the number of frames of the unit are stored as unit-level acoustic feature representation. 
  
  Three different RAEs have been trained using different set of features.
  \begin{itemize}
    \item MGC static features alone (x=50). 
    \item MGC with delta features (x=150). 
    \item MGC and BAP with delta features (x=228) . 
  \end{itemize}
  The input, output weights were initialized with Gaussian distribution with variance 0.01 and 0.01 respectively for MGC and 0.003 and 0.003 for MGC with deltas networks. We found that the network was sensitive to scale of variance at input and output to converge to a good local minimum.
  
  During synthesis, the RBN features are given as input to the decoder and the output sequence is predicted. Note here that while training the RAE the past output is taken from the ground truth parameter sequence during testing time the predicted past output is used.  
     
  The Mel-cepstral distortion between the original and re-synthesized spectra is reported in Table \ref{tab:AE_MCD}. It can be seen that MGC features either static or including the dynamic features perform well. Usually MCD of less than 4dB is an indicator of high-quality synthesis \cite{mcd}. The third row in the table (MGC-DD-nml) refers to experiments performed by normalizing the MGC-DD features between 0.01 and 0.99. The normalization was not needed for MGC static coefficients but was helpful when considering the dynamic coefficients. This may be because the dynamic range of static and dynamic coefficients are very different. However when BAP coefficients are appended the RAE did not perform well although normalization was done. This suggest that training different networks for BAP and MGC may be more appropriate. The MOS scores of synthesized speech signals and the natural speech are depicted in column 1 of Table \ref{tab:AE_MCD} which shows that the quality of the AbS speech is high. An example spectrogram of natural speech recording and the proposed RBN feature based AbS is shown in Fig. \ref{fig:RBNAbS_sgram}
 
\begin{table}[th]
    \caption{Mel-cepstral distortion (MCD) and Mean Opinion Score (MOS) for RBN AbS \label{tab:AE_MCD} }
    \vspace{2mm}
    \centerline{
    \begin{tabular}{|c|c|c|}
    \hline
    Feature & MOS & MCD \\
    \hline
    MGC        & $4.04$   & $1.49$~~~ \\
    MGC-DD     & $3.67$   & $2.18$~~~ \\
    MGC-DD-nml & $3.65$   & $1.96$~~~ \\
    MGC+BAP-DD & $2.58$   & $3.17$~~~ \\
    \hline
    \end{tabular}
}
\end{table}

\subsection{SPSS using RBN feature}
In this section, experiments using RBN as acoustic features in a DNN based SPSS system is explained. As a baseline system we take frame-level text and acoustic feature sequences 2800000, 70000 and 70000 feature vectors respectively for training, validation and testing. A 345L 1000R 1000R 50L system was trained to predict frame level acoustic parameters (MGCs). For the proposed system, RBN features corresponding to 200000 units are used for training another DNN. 2800000 feature vectors are reduced to about 200000 feature vectors when mapping at the unit-level instead of frame level a reduction factor of 14. The architecture of the DNN was 345L 1000R 1000R 500L. The predicted RBNs are then fed to the decoder network of the trained RAE as discussed in the previous section. The RAE trained with static MGC was used since it was the best performing amongst the AbS systems. The output speech is synthesized using the predicted MGC, natural BAP and $f_0$ using STRAIGHT vocoder \cite{STRAIGHT}. Note that dynamics were not used during training of any of the systems and hence MLPG \cite{mlpg} could not be applied which results in discontinuous synthesis. This probably is the reason for low MOS score in our results.

The MCD and MOS scores of the baseline and the proposed are reported in table \ref{tab:SPSS_MCD}. It can be seen that both the SPSS systems give rise to similar MCD and MOS scores. 

The code for replicating the experiments can be found online 
\footnote{https://goo.gl/kJBGyg}.

\begin{table}[th]
    \caption{Mel-cepstral distortion for SPSS systems \label{tab:SPSS_MCD} }
    \vspace{2mm}
    \centerline{
    \begin{tabular}{|c|c|c|}
    \hline
    System & MOS & MCD \\
    \hline
    Baseline &$2.7$ & $5.6$~~~ \\
    Proposed &$2.5$ & $5.8$~~~ \\
    \hline
    \end{tabular}
}
\end{table}

  \section{Conclusions and Scope For Future Work}
  \label{sec:con}
  In this work, we have explored the possibility of compactly representing the acoustic parametric sequences of units with a single vector using sequence-to-sequence auto-encoders. While achieving compression it was shown that this process of recurrent auto-encoding does not affect the speech quality. The intermediately obtained RBN features were used as unit-level acoustic features which were then mapped from the text features. This representation was shown to greatly reduce the computational cost of text-to-speech mapping. A DNN based SPSS system was built to demonstrate that the RBN features can indeed be used as acoustic features within SPSS framework.
  
  There are many interesting extensions to the current work. At feature-level auto encoding can be done at waveform level instead of high level features like spectrum used in this work. The use of RAE as a post filter is one other direction for exploration. Since the phoneme durations are known apriori during the synthesis the above training method fits into the post-filtering scheme easily. 
  
  \section{Acknowledgements}  
    The authors like to acknowledge TCS for partially funding first authors PhD. Also authors would like to thank speech and vision lab members for participating in the listening tests.

  \newpage
  \eightpt
  \bibliographystyle{IEEEtran}

  \bibliography{mybib_tejas}


\end{document}